\documentclass[a4paper]{jpconf}
\usepackage{graphicx}
\usepackage{color}
\newcommand{\mincir}{\raise
-3.truept\hbox{\rlap{\hbox{$\sim$}}\raise4.truept\hbox{$<$}\ }}
\newcommand{\magcir}{\raise
-3.truept\hbox{\rlap{\hbox{$\sim$}}\raise4.truept\hbox{$>$}\ }}
\newcommand{\minmag}{\raise
-3.truept\hbox{\rlap{\hbox{$<$}}\raise5.truept\hbox{$<$}\ }}
\newcommand{\be}{\begin{equation}}
\newcommand{\ee}{\end{equation}}
\newcommand{\ba}{\begin{eqnarray}}
\newcommand{\ea}{\end{eqnarray}}

\begin{document}

\title{The Dark Energy Equation of State using
Alternative Cosmic High-$z$ Tracers}

\author{M. Plionis${^{1,2}}$, R. Terlevich$^{2}$, S. Basilakos$^3$,
  F. Bresolin$^4$, E. Terlevich$^2$, J. Melnick$^5$, R. Chavez$^2$}

\address{$^1$ Institute of Astronomy \& Astrophysics, National Observatory of Athens,
    Palaia Penteli 152 36, Athens, Greece.\\
    $^2$ Instituto Nacional de Astrof\'{\i}sica Optica y Electr\'onica, AP 51
    y 216, 72000, Puebla, M\'exico.\\
    $^3$Academy of Athens, Research Center for Astronomy and Applied Mathematics,
 Soranou Efesiou 4, 11527, Athens, Greece \\
    $^4$ Institute for Astronomy of the University of Hawaii, 2680
    Woodlawn Drive, 96822 Honolulu, HI USA \\
    $^5$ European Southern Observatory, Alonso de Cordova 3107, Santiago, Chile
}

\ead{mplionis@astro.noa.gr}

\begin{abstract}
We propose to use alternative cosmic tracers to measure the dark
energy equation of 
state and the matter content of the Universe [w$(z)$ \& $\Omega_m$]. Our 
proposed method consists of two components: (a) tracing the 
Hubble relation using HII galaxies which can be detected 
up to very large redshifts, $z\sim 4$, as an 
alternative to supernovae type Ia, and (b) measuring the 
clustering pattern of X-ray selected AGN at a median 
redshift of $\sim 1$. Each component of the method can in itself 
provide interesting constraints on the cosmological parameters, 
especially under our anticipation that we will reduce the 
corresponding random and systematic errors significantly. 
However, by joining their likelihood functions we will be 
able to put stringent cosmological constraints and break 
the known degeneracies between the {\em dark energy} equation 
of state (whether it is constant or variable) and the matter 
content of the universe and provide a powerful and alternative 
route to measure the contribution to the global dynamics and 
the equation of state of {\em dark energy}. 
A preliminary joint analysis of X-ray selected AGN clustering (based on the
largest to-date XMM survey; the 2XMM) and the currently largest SNIa
sample, the {\em Constitution} set (Hicken et al.), 
using as priors a flat universe and the WMAP5 normalization
of the power-spectrum, provides: $\Omega_{\rm m}=0.27\pm 0.02$ and
w$=-0.96\pm 0.07$. 
Equivalent and consistent results are provided by the joint analysis of X-ray selected AGN
clustering and the latest Baryonic Acoustic Oscillation measures, providing:
$\Omega_{\rm m}=0.27\pm 0.02$ and w$=-0.97\pm 0.04$. 
\end{abstract}

\section{Introduction}
We live in a very exciting period for our understanding of the
Cosmos. Over the past decade the accumulation and detailed analyses of
high quality cosmological data (eg., supernovae type Ia, CMB
temperature fluctuations, galaxy clustering, high-z clusters of
galaxies, etc.) have strongly suggested that we live in a 
flat and accelerating universe, which contains at least some sort of
cold dark matter to explain the clustering of extragalactic sources,
and an extra component which acts as having a negative pressure, as for
example the energy of the vacuum (or in a more general setting the so
called {\em dark energy}), to explain the observed accelerated cosmic
expansion  (eg. Riess, et al. 1998; 2004; 2007, Perlmutter et
al. 1999; Spergel et al. 2003, 2007, Tonry et al. 2003; Schuecker et
al. 2003; Tegmark et al. 2004; Seljak et al. 2004; Allen et al. 2004;
Basilakos \& Plionis 2005; 2006; 2009; Blake et al. 2007; Wood-Vasey et
al. 2007, Davis et al. 2007; Kowalski et al. 2008, Komatsu et
al. 2008; Hicken et al. 2009, etc).

Due to the absence of a well-motivated fundamental theory, there have
been many theoretical speculations regarding the nature of the exotic
{\em dark energy}, on whether it is a cosmological constant, a scalar or
vector fields which provide a time varying dark-energy equation of
state, usually parametrized by: 
\begin{equation}
p_Q= {\rm w}(z) \rho_Q\;,
\end{equation} 
with $p_Q$ and $\rho_Q$ the pressure and density of the exotic dark
energy fluid and 
\be 
{\rm w}(z)={\rm w}_0 + {\rm w}_1 f(z) \;,
\label{eqstatez}
\ee 
with w$_0=$w$(0)$ and $f(z)$ an increasing function of redshift [ eg.,
$f(z)=z/(1+z)$] (see
Peebles \& Ratra 2003 and references therein, Chevalier \& Polarski
2001, Linder 2003, Dicus \& Repko 2004; Wang \& Mukherjee 2006). Of
course, the equation of state could be such that w does not evolve
cosmologically. 

Two very extensive recent reports have identified {\em dark energy} as a top
priority for future research: "Report of the Dark Energy Task Force 
(advising DOE, NASA and NSF) by Albrecht et al. (2006), and ``Report of
the ESA/ESO Working Group on Fundamental Cosmology'', by Peacock et
al. (2006).  It is clear that one of the most important questions in
Cosmology and cosmic structure formation is related to the nature of
{\em dark energy} (as well as whether it is the sole interpretation of the
observed accelerated expansion of the Universe) and its interpretation
within a fundamental physical theory. To this end a large number of
very expensive experiments are planned and are at various stages of
development, among which the {\em Dark Energy Survey} 
(DES: {\tt http://www.darkenergysurvey.org/}), the {\em Joint Dark Energy
  Mission} (JDEM: {\tt http://jdem.gsfc.nasa.gov/}), {\em HETDEX} 
({\tt http://www.as.utexas.edu/hetdex/}), Pan-STARRS: {\tt http://pan-starrs.ifa.hawaii.edu}, etc.

Therefore, the paramount importance of the detection and
quantification of {\em dark energy} for our understanding of the cosmos and
for fundamental theories implies that the results of the different
experiments should not only be scrutinized, but alternative, even
higher-risk, methods to measure {\em dark energy} should be developed and
applied as well.
 It is within this paradigm that our current work falls.
Indeed, we wish to constrain the {\em dark energy} equation of state
using, individually and in 
combination, the Hubble relation and large-scale structure (clustering) methods, but
utilizing alternative cosmic tracers for both of these components.
 
From one side we wish to trace the Hubble function using HII 
galaxies, which can be observed at higher redshifts than those
sampled by current SNIa surveys and thus at distances where the Hubble function
is more sensitive to the cosmological parameters. The HII galaxies can
be used as standard candles (Melnick, Terlevich \& Terlevich 2000,
Melnick 2003; Siegel et al. 2005; Plionis et al. 2009) 
due to the correlation between their velocity
dispersion, metallicity and $H_{\beta}$ luminosity (Melnick 1978, Terlevich \&
Melnick 1981, Melnick, Terlevich \& Moles 1988), once we reduce
significantly their distance modulus uncertainties, which at present are
unacceptably large for precision cosmology ($\sigma_\mu\simeq 0.52$
mag; Melnick, Terlevich \& Terlevich 2000).
Furthermore, the use
of such an alternative high-$z$ tracer will enable us to check the SNIa based results
and lift any doubts that arise from the fact that they are the only
tracers of the Hubble relation used to-date (for possible usage of GRBs
see for example, Ghirlanda et al. 2006; Basilakos \& Perivolaropoulos
2008)\footnote{
GRBs appear to be anything but standard candles, having a very wide
range of isotropic equivalent luminosities and energy outputs.
Nevertheless, correlations between various properties of
the prompt emission and in some cases also the afterglow emission
have been used to determine their distances.
A serious problem that hampers a straight forward use of GRBs as
Cosmological probes is the intrinsic
faintness of the nearby events, a fact which introduces a bias towards low (or high)
values of GRB observables and therefore the extrapolation
of their correlations to low-$z$ events is faced with
serious problems.
One might also expect a significant evolution of the
intrinsic properties of GRBs with redshift (also between
intermediate and high redshifts) which can be hard to disentangle
from cosmological effects. Finally, even if a reliable scaling relation
can be identified and used, the scatter in the resulting
luminosity and thus distance modulus is still fairly large.}.

From the other side we wish to use X-ray selected AGN at a median
redshift of $\sim 1$, which is roughly the
peak of their redshift distribution (see Basilakos et al. 2004; 2005,
Miyaji et al. 2007), in order to determine their clustering pattern
and compare it with that predicted by different cosmological models. 

Although each of the previously discussed components of our project
(Hubble relation using HII galaxies and angular/spatial
clustering of X-ray AGN) will provide interesting and relatively
stringent constraints on the cosmological parameters, especially under
our anticipation that we will reduce significantly the corresponding
random and systematic errors, it is the combined likelihood of these
two type of analyses that enables us to break the known
degeneracies between cosmological parameters and determine with great
accuracy the {\em dark energy} equation of state (see 
Basilakos \& Plionis 2005; 2006; 2009).

Below we present the basic methodology of each of the two main components of
our proposal, necessary in order to constrain the {\em dark energy}
equation of state.

\section{Cosmological Parameters from the Hubble Relation}
It is well known that in the matter dominated epoch and in flat
universes, the Hubble relation depends on the cosmological
parameters via the following equation:
\be
H(z) = H_{0} E(z) \;\;\;\; {\rm with} \;\;\; 
E(z)=\left[\Omega_m (1+z)^3 + \Omega_Q \exp \left(3 \int_0^z
  \frac{1+{\rm w}(x)}{1+x} {\rm d}x\right) \right]^{\frac{1}{2}}\;,
\ee
which is simply derived from Friedman's equation. We remind the reader
that $\Omega_m$ and $\Omega_Q (\equiv 1-\Omega_{m})$ are the present fractional contributions to the
total cosmic mass-energy density of the matter and dark energy source
terms, respectively.

Supernovae SNIa are considered standard candles at peak luminosity
and therefore they have been used not only to determine the Hubble constant (at
relatively low redshifts) but also to trace the curvature of the
Hubble relation at high redshifts (see Riess et al. 1998, 2004, 2007; Perlmutter et
al. 1998, 1999; Tonry et al. 2003; Astier et al. 2006; Wood-Vasey et
al. 2007; Davis et al. 2007; Kowalski et al 2008; Hicken et al. 2009). 
Practically one relates the distance modulus of the SNIa to its luminosity
distance, $d_L$, through which the cosmological parameters enter:
\be
\mu = m-M = 5 \log d_L + 25 \;\;\;\; {\rm where} \;\;\;\;
d_{L} =  (1+z) \frac{c}{H_0} \int_0^z \frac{{\rm d}z}{E(z)} \;.
\ee
The main result of numerous studies using this procedure is that distant 
SNIa's are dimmer on average by $\sim$0.2 mag than what expected 
in an Einstein-deSitter model, which translates in them being $\sim 10\%$ further
away than expected.

The amazing consequence of these results is that 
we live in an accelerating phase
of the expansion of the Universe, an assertion that needs to be
scrutinized on all possible levels, one of which is to verify the accelerated expansion of the
Universe using alternative to SNIa's extragalactic standard candles. Furthermore, the cause and rate of the
acceleration is of paramount importance, ie., the {\em dark energy} equation of
state is the next fundamental item to search for 
 and to these directions we hope to contribute with our current project.

\subsection{Theoretical Expectations:}
To appreciate the magnitude of the Hubble relation variations 
due to the different {\em dark energy} equations of state, we plot in Figure 1 the relative
deviations of the distance modulus, $\Delta\mu$, of different {\em dark-energy}
models from a nominal {\em standard} (w$=-1$) 
$\Lambda$-cosmology (with $\Omega_m=0.27$ and
$\Omega_{\Lambda}=0.73$), with the relative deviations defined as:
\be
\Delta\mu=\mu_{\Lambda} - \mu_{\rm model} \;.
\ee
The parameters of the different models used are shown in Figure
1. As far as the {\em dark-energy} equation of state parameter is
concerned, we present the deviations from the {\em standard} model 
of two models with a constant w value and of two models with an
evolving equation of state parameter, utilizing the form of eq.(\ref{eqstatez}).
 In the left panel of Figure 1 we present
results for selected values of $\Omega_m$, while in the right panel
we use the same {\em dark-energy} equations of state parameters but
for the same value of $\Omega_m (=0.27)$ (ie., we eliminate the
$\Omega_m-{\rm w}(z)$ degeneracy).
\begin{figure}
\centering
\resizebox{14cm}{7.5cm}{\includegraphics{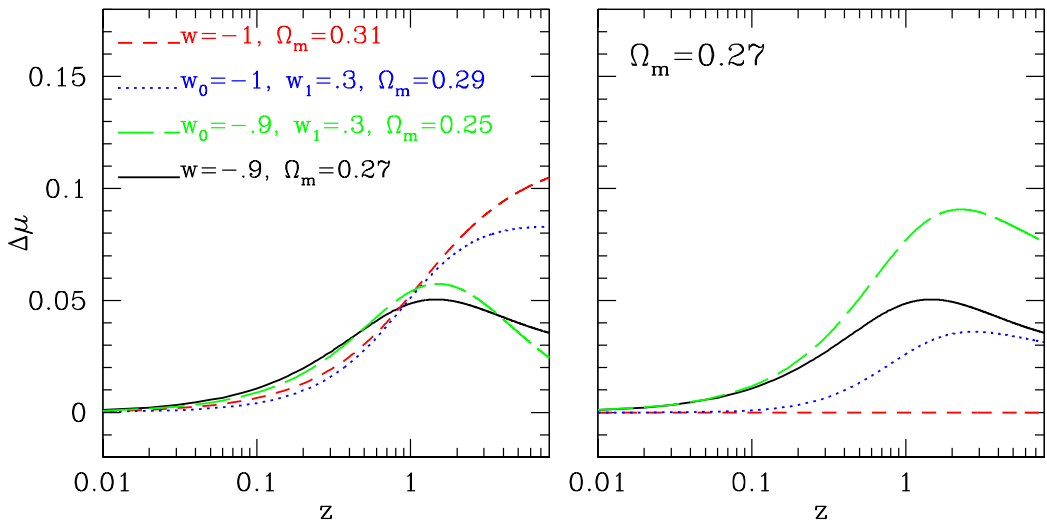}}
\caption{{\em Left Panel:} The expected distance modulus difference between the
  {\em dark-energy} models shown and the reference
  $\Lambda$-model (w$=-1$) with $\Omega_{m}=0.27$. {\em Right Panel:}
  The expected distance modulus differences once the $\Omega_m$-w$(z)$
  degeneracy is broken (imposing the same $\Omega_m$ value as in the comparison model).}
\end{figure} 

Three important observations should be made from Figure 1:
\begin{enumerate}
\item The relative magnitude deviations between the different {\em dark-energy}
  models are quite small (typically $\mincir 0.1$ mag), which puts severe pressure on
  the necessary photometric accuracy of the relevant observations.
\item The largest relative deviations of the distance moduli occur at
redshifts $z\magcir 1.5$, and thus at quite larger redshifts than those
currently traced by SN Ia samples, and
\item There are strong degeneracies between the different cosmological
  models at redshifts $z\mincir 1$, but in some occasions even up to
  much higher redshifts (one such example is shown in Figure 1 between
  the models with $(\Omega_m, {\rm w}_0, {\rm w}_1)=(0.31,-1,0)$ and 
$(0.29,-1,0.3)$.
\end{enumerate}
Luckily, such degeneracies can be broken, as discussed
already in the introduction, by using other cosmological tests
(eg. the clustering of extragalactic sources, the CMB shift parameter, BAO's, etc). Indeed,
current evidence overwhelmingly show that the total matter content of the universe
is within the range: $0.2\mincir \Omega_m \mincir 0.3$, a fact that reduces significantly
the degeneracies between the cosmological parameters.

\subsection{Larger numbers or higher redshifts ?}
In order to define an efficient strategy to put stringent constraints
on the {\em dark-energy} equation of state, we have decided to re-analyse two recently compiled
SNIa samples, the Davis et al. (2007) [hereafter {\em D07}] compilation of 192 SNIa
(based on data from Wood-Vasey et al. 2007, Riess et al. 2007 and
Astier et al. 2007) and the
{\em Constitution} compilation of 397 SNIa (Hicken et al. 2009).
Note that the two
samples are not independent since most of the {\em D07} is included in
the {\em Constitution} sample.

Firstly, we present in the left panel of Figure 2 the {\em Constitution} SNIa distance moduli
overploted (red-line) with the theoretical expectation of a flat cosmology
with $(\Omega_m, {\rm w})=(0.27,-1)$. In the right panel we
plot the distance moduli difference between the SNIa data and the previously
mentioned model. To appreciate the level of accuracy needed in order
to put constraints on the equation of state parameter, we also plot the
distance moduli difference between the reference $(\Omega_m, {\rm
  w})=(0.27,-1)$ and the $(\Omega_m, {\rm w})=(0.27,-0.85)$ models (thin blue line).

\begin{figure}
\centering
\resizebox{16cm}{7.5cm}{\includegraphics{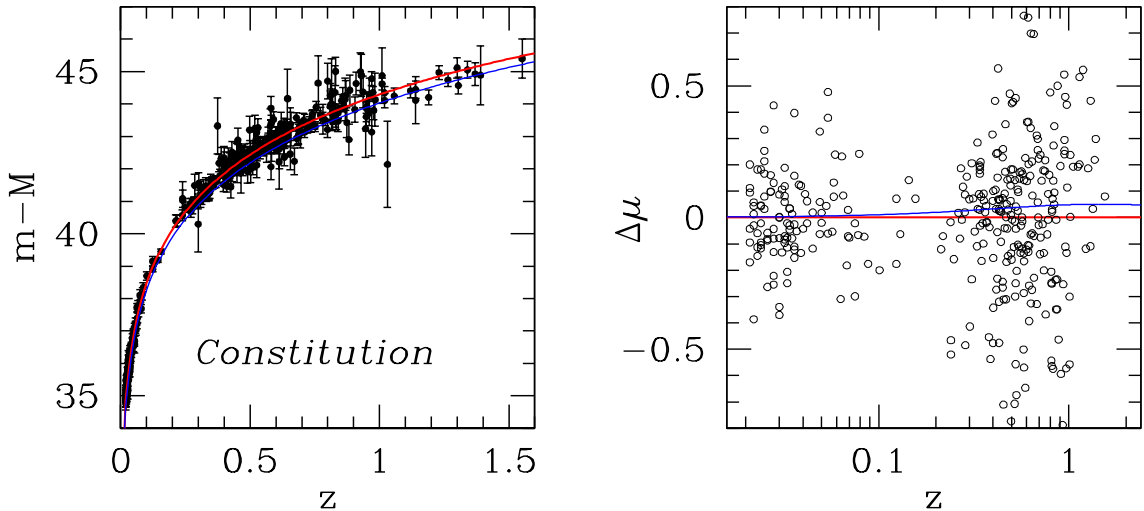}}
\caption{{\em Left Panel:} SNIa distance moduli as a function of redshift. {\em
    Right Panel:} Distance moduli difference between the
  $\Lambda$-model and the SNIa data. The blue line is the
  corresponding difference between the reference (w$=-1$) and the
  w$=-0.85$ {\em dark-energy} models.}
\end{figure}

We proceed to analyse the SNIa data by defining the usual likelihood
estimator\footnote{Likelihoods
  are normalized to their maximum values.} as:
\be
{\cal L}^{\rm SNIa}({\bf p})\propto {\rm exp}[-\chi^{2}_{\rm
  SNIa}({\bf p})/2]
\ee
where ${\bf p}$ is a vector containing the cosmological 
parameters that we want to fit for, and
\be
\chi^{2}_{\rm SNIa}({\bf p})=\sum_{i=1}^{N} \left[ \frac{ \mu^{\rm th}(z_{i},{\bf p})-\mu^{\rm obs}(z_{i}) }
{\sigma_{i}} \right]^{2} \;\;,
\ee 
where $\mu^{\rm th}$ is given by eq.(3), $z_{i}$ is the observed redshift and $\sigma_{i}$ the observed
distance modulus uncertainty. 
Here we will constrain our analysis within the framework 
of a flat ($\Omega_{\rm tot}=1$) cosmology and therefore 
the corresponding vector ${\bf p}$ is: ${\bf p}\equiv (\Omega_m, {\rm
  w}_0, {\rm w}_1)$. We will use only SNIa with $z>0.02$ in order to avoid
redshift uncertainties due to the local bulk flow (eg. Hudson et
al. 1999 and references therein).

\subsubsection{\em Larger Numbers?} 
The first issue that we wish to address is how better have we done in
imposing cosmological constraints by increasing the available SNIa
sample from 181 to 366
, ie., increasing the
sample by more than 100\%. In Table 1 we present various solutions
using each of the two previously mentioned samples 
Note that since
only the relative distances of the SNIa are accurate and not their
absolute local calibration, we always marginalize with respect to the
internally derived Hubble constant (note that fitting procedures exist
which do not need to {\em a priori} marginalize over the
internally estimated Hubble constant; eg., Wei 2008). Although the
derived cosmological parameters are consistent between the two data
sets, possibly indicating the robustness of the method, the corresponding
goodness of fit (the reduced $\chi^2$) is significantly 
larger in the case of the {\em Constitution} set (1.21 compared to 1.045 of
the {\em D07} set). 
This appears to be the outcome of the different approaches chosen in
order to join the different contributing SNIa sets. According to
Hicken 2009 (private communication) in the case of the {\em D07} the
nearby SNIa were imposed to provide a $\chi^2/df\simeq 1$ by hand,
while no such fine-tuning was imposed on the UNION set (on which the
{\em Constitution} set is based). A secondary reason could be that 
the latter set includes
distant SNIa which have typically larger distance modulus uncertainties,
with respect to those used in {\em D07}.
Overall, the higher $\chi^2/df$ value of the {\em Constitution} set
should be attributed to a typically lower uncertainty in $\mu$.
As a crude test, we have increased by 20\% the distance modulus
uncertainty of the {\em Constitution} nearby SNIa ($z\mincir 0.4$) 
and indeed we obtain $\chi^2/df\simeq 1.07$, similar to that of {\em D07}.

\begin{figure}
\centering
\label{main_snia}
\resizebox{15cm}{7.3cm}{\includegraphics{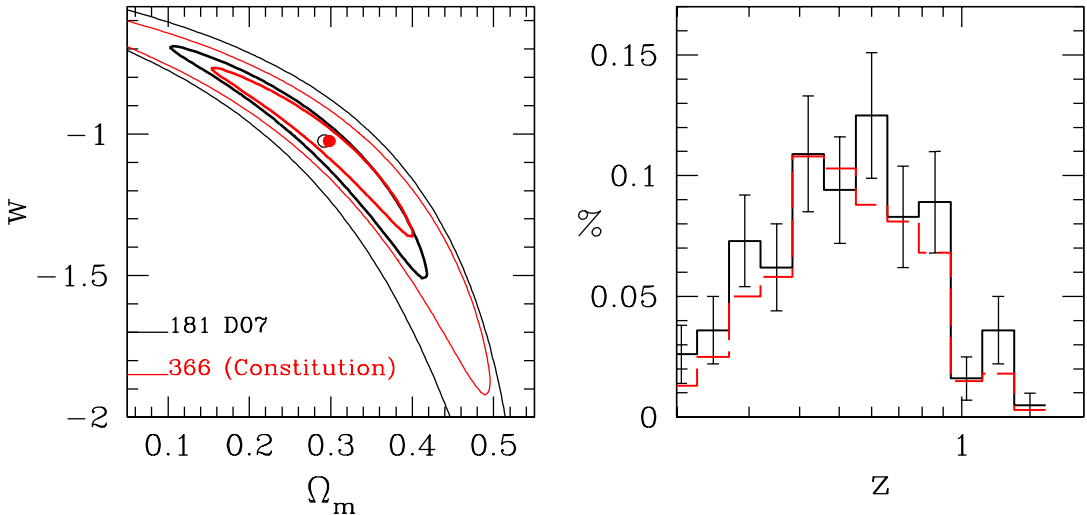}}
\caption{ {\em Left panel:} Cosmological parameter solution space 
  using either of the two SNIa data sets ({\em Constitution}: red
  contours and {\em D07}: black contours). Contours corresponding to the 1 and 3$\sigma$ confidence levels are
shown (ie., plotted where $-2{\rm ln}{\cal L}/{\cal L}_{\rm max}$ is equal
to 2.30 and 11.83, respectively).
{\em Right Panel:} Normalized redshift distributions of the two SNIa data
sets.}
\end{figure} 

\begin{figure}
\centering
\label{dev}
\resizebox{15cm}{7cm}{\includegraphics{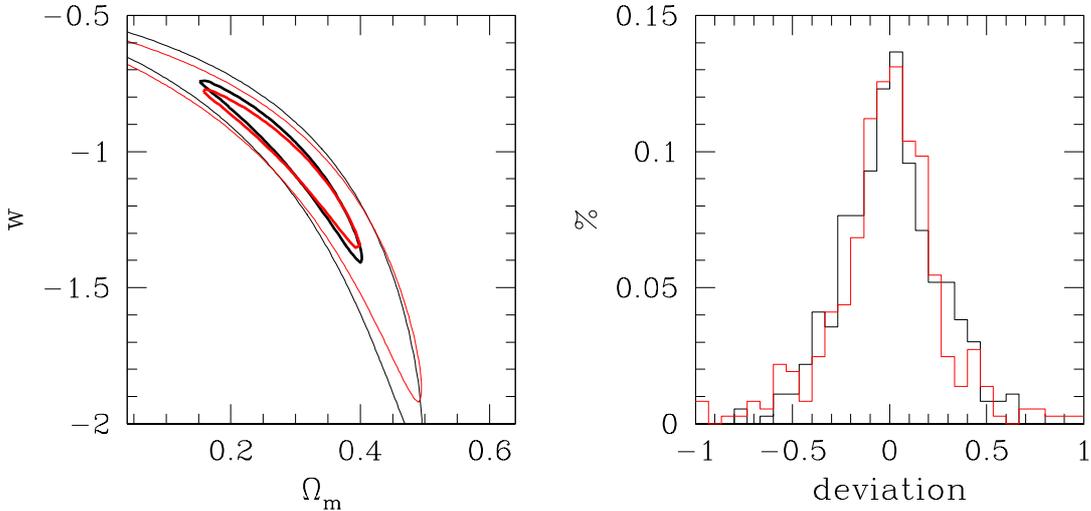}}
\caption{{\em Left Panel:} Comparison between the {\em Constitution} SNIa
  constraints (red contours) and those derived by a Monte-Carlo 
procedure designed to closely
  reproduce them (for clarity we show only contours corresponding to 1
  and 3 $\sigma$ confidence levels). {\em Right Panel:} The {\em
    Constitution} SNIa distance modulus
  deviations from the best fit model $(\Omega_m, {\rm w})\simeq
  (0.30,-1.01)$; see Table 1 -
  and a random realization of the model deviations (red histogram).}
\end{figure} 

In Figure 3 we can also see that although the SNIa sample has doubled in 
size, the well-known {\em banana}
shape region of the ($\Omega_m,$w) solution space, indicating the
degeneracy between the two cosmological parameters, is reproduced by
both data sets. However, there is a reduction of the size of
the solution space when using the {\em Constitution} SNIa
compilation (see also Table 1), and at roughly the level expected from
Poisson statistics.

A first conclusion is therefore that {\em the increase by $\sim 100\%$
  of the Constitution sample has not
provided significantly more stringent constraints to the cosmological
parameters.} 
We have verified that the larger number of SNIa's in the {\em Constitution} sample are
not preferentially located at low-$z$'s (see right panel of Fig.2) -
in which case we should 
have not expected more stringent
cosmological constraints using the latter SNIa sample, but rather
have a very similar $z$-distribution.

\begin{table}
\caption{\small Cosmological parameter fits using the SNIa data within
  flat cosmologies. Note that for
  the case where ${\bf p}=(\Omega_m, {\rm w})$ (ie., last row), the errors
  shown are estimated after marginalizing with respect to the other
  fitted parameter.}
\tabcolsep 8pt

\begin{tabular}{|ccc|ccc|} \hline
  \multicolumn{3}{c}{\em D07}  
&\multicolumn{3}{c}{\em Constitution} \\ \hline 
  w & $\Omega_{m}$ & $\chi^2_{\rm min}$/df 
& w & $\Omega_{m}$ & $\chi^2_{\rm min}$/df \\ \hline
${\bf -1}$ & $0.280^{+0.025}_{-0.015}$ & 187.03/180 
&${\bf -1}$ & $0.286^{+0.012}_{-0.018}$ &  439.78/365 \\
$-1.025^{+0.060}_{-0.045}$ & $0.292\pm 0.018$ & 187.02/179 
& $-1.025\pm 0.030$ & $0.298\pm 0.012$ & 439.79/364
\\ \hline
\end{tabular}

\end{table}

We already have a strong hint, from the previously
presented comparison between the {\em D07} and {\em Constitution}
results, that increasing the number of Hubble
relation tracers, covering the same redshift range and with the current
level of uncertainties as the available SNIa samples, 
does not appears to be an effective avenue for providing further
constraints of the cosmological parameters. 

\subsubsection{\em Lower uncertainties or higher-$z$'s:}
We now resort to a Monte-Carlo procedure which will help us 
investigate which of the following two directions, which bracket many
different possibilities, would provide more stringent cosmological constraints:

\begin{itemize}
\item Reduce significantly the distance modulus uncertainties of SNIa, 
  tracing however the same redshift range as the currently available samples, or
\item use tracers of the Hubble relation located at 
  redshifts where the models show their largest relative differences
  ($z \magcir 2$), with distance modulus uncertainties comparable to that of
  the highest redshift SNIa's ($\langle \sigma_{\mu}\rangle \simeq 0.4$)
\end{itemize}

The Monte-Carlo procedure is based on using the observed high-$z$ SNIa
distance modulus uncertainty 
distribution ($\sigma_\mu$) and a model to assign random $\mu$-deviations from a
reference $H(z)$ function, that reproduces exactly the original
banana-shaped contours of the $(\Omega_m, {\rm w})$ solution space of Figure
3 (left panel). Indeed, after a trial and error procedure we have
found that
by assigning to each SNIa (using their true redshift) a distance
modulus deviation ($\delta \mu$) from a reference model
having a Gaussian distribution with zero mean
and variance given by the observed $\langle \sigma_\mu \rangle^2$, 
and using as the relevant individual
distance modulus uncertainty
the following: $\sigma^2_i=\sqrt{(1.2
\delta\mu_i)^2 + \phi^2}$ (with $\phi$ a random Poisson
deviate within $[-0.01, 0.01]$) we reproduce exactly the banana-shaped
solution range of the reference model.
This can be seen clearly in the
upper left panel of Figure 4, where we plot the original {\em
  Constitution} SNIa
solution space (red contours) and the model solution space (black contours).
In the right panel we show the distribution of the true SNIa
deviations from the best fitted model as well as a random
realization of the model deviations.
\begin{figure}
\centering
\label{comp_m}
\resizebox{14cm}{7cm}{\includegraphics{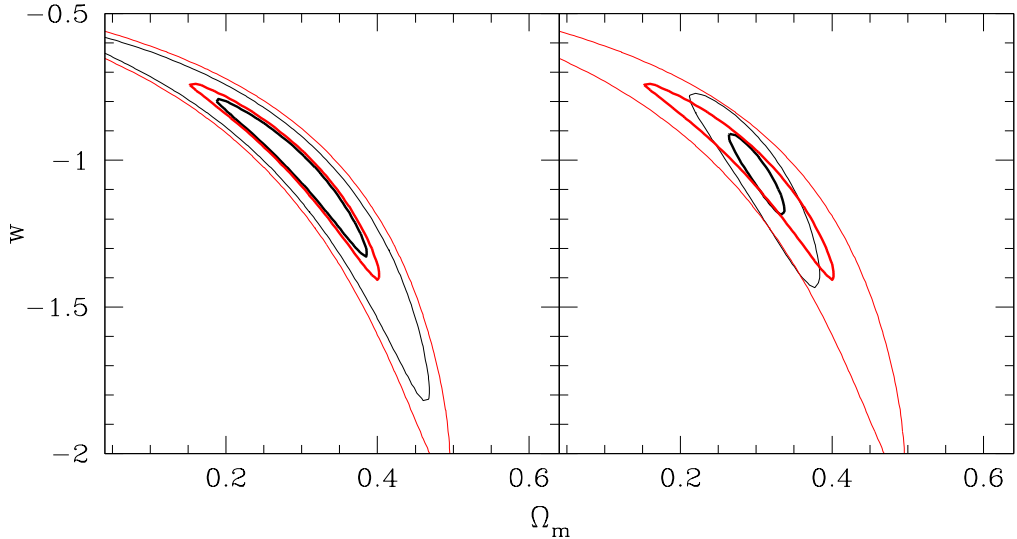}}
\caption{Comparison of the model {\em Constitution} SNIa 
  constraints (red contours) with 
those derived by reducing to half
  their uncertainties ({\em left panel}, and with those derived by adding a sample of
 82 high-$z$ tracers ($2.7\mincir z \mincir 3.5$) with a
 distance modulus mean uncertainty of $\sigma_\mu\simeq 0.38$ ({\em
   right panel}. For clarity we show only contours corresponding to the 1
  and 3 $\sigma$ confidence levels.}
\end{figure} 

Armed with the above procedure we can now address the questions posed
previously. Firstly, we reduce to half the random deviations of the
SNIa distance moduli from the
reference model (with the corresponding reduction of the relevant
uncertainty, $\sigma_i$). The results of the likelihood analysis can be seen in the
left panel of Figure 5. There is a reduction of the range of
the solution space, but indeed quite a small one. 
Secondly, we add to the {\em Constitution} SNIa sample, a mock
high-$z$ subsample of 82 objects constructed as follows: For each
$z>0.68$ SNIa we add one mock SNIa having as a redshift $z+\delta z$
where $\delta z=2$ and the same distance modulus uncertainty.
Note that these new 82 SNIa are distributed between
$2.68\mincir z\mincir 3.55$, ie., in a range where the largest
deviations between the different cosmological models occur (see Figure 1).
The deviations from the reference model of
these additional SNIa are based on their original $\mu$ uncertainty 
 distribution (ie., we have assumed that the new high-$z$ tracers will
 have similar uncertainties as their $z\magcir 0.68$ counterparts,
 which on average is $\langle \sigma_\mu\rangle \simeq 0.38$).
We now find a significantly reduced solution space (right panel
of Figure 5), which shows that indeed by increasing the $H(z)$
tracers by a few tens, at those redshifts where the largest deviations
between models occur, can have a significant impact on the recovered
cosmological parameter solution space. 
Similarly, if we use a reference model with an evolving dark-energy equation of state 
(as that of eq. 2), and after marginalizing with respect to $\Omega_m$, we also find 
a significantly larger reduction of the (w$_0$, w$_1$) solution space when we include 
the high redshift tracer subsample (Fig.6, right panel) with respect to the
distance modulus uncertainty reduction (by 1/2) case (Fig. 6, left panel). 

The main conclusion of the previous analysis is that a relatively
better strategy
to decrease the uncertainties of the cosmological parameters, based on
the Hubble relation, is to use standard candles which trace also the redshift
range $2\mincir z \mincir 4$. 
Below we present one such possibility by suggesting an alternative, to the
SNIa standard candles, namely HII galaxies (eg. Melnick
2003; Siegel et al. 2005).

\begin{figure}
\centering
\label{comp_m}
\resizebox{14cm}{7cm}{\includegraphics{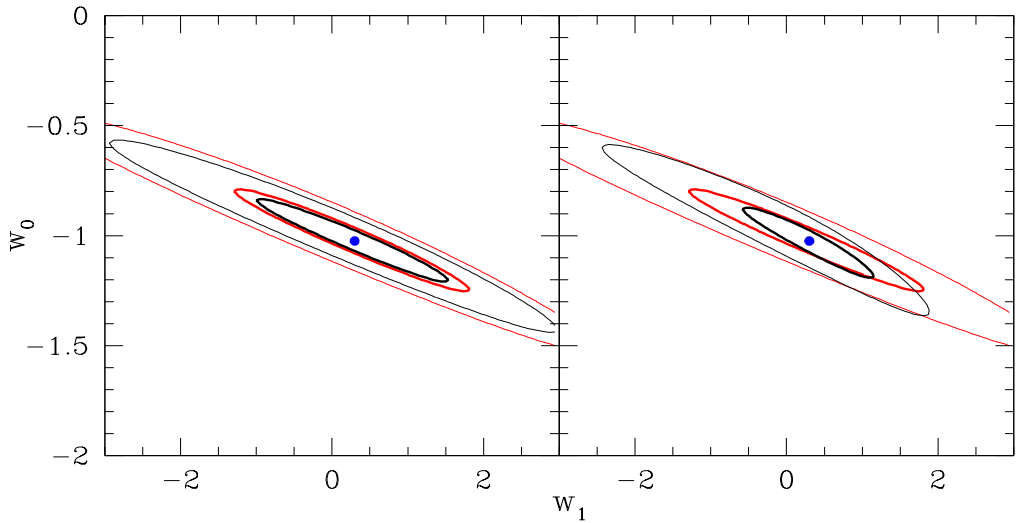}}
\caption{Similar as in Figure 5, but allowing for an evolving DE equation of state according to eq. (2)
and after 
marginalizing with respect to $\Omega_m$. The input cosmological model has (w$_0$, w$_1)=(-1.025, 0.3)$ 
and is represented by the red contours.
Again we show for clarity only contours corresponding to the 1
  and 3 $\sigma$ confidence levels.}
\end{figure}

\subsection{ Hubble Relation using HII galaxies}
We now reach our suggestion to use an alternative and potentially 
very powerful technique to estimate cosmological
distances, which is the relation between the luminosity of the $H_{\beta}$ line and
the stellar velocity dispersion, measured from the line-widths, of HII regions
and galaxies (Terlevich \& Melnick 1981, Melnick, Terlevich \& Moles
1988). The cosmological use of this distance indicator has been tested
in Melnick, Terlevich \& Terlevich (2000) and Siegel et al (2005) (see
also the review by Melnick 2003).
The presence of O and B-type stars in HII regions 
causes the strong Balmer line emission, in both $H_{\alpha}$ and $H_{\beta}$. Furthermore,
the fact that the bolometric luminosities of HII galaxies are
dominated by the starburst component implies that their luminosity per
unit mass is very large, despite the fact that the galaxies are
low-mass. Therefore they can be observed at very large redshifts, and
this fact makes them cosmologically very interesting
objects. Furthermore, it has been shown that the $L(H_{\beta})-\sigma$ correlation
holds at large redshifts (Koo et al. 1996, Pettini et al. 2001, Erb et
al. 2003) and therefore it can be used to trace the Hubble relation
at cosmologically interesting distances. One of the most important
prerequisites in using such relations, as distance estimators, is the
accurate determination of their zero-point. To this end, Melnick et al
(1988) used giant HII regions in nearby late-type galaxies and derived
the following empirical relation (using a Hubble constant of $H_0=71$
km/sec/Mpc):

\begin{equation}
\log L(H_{\beta}) = \log M_z + 29.60 \;\; {\rm with} \;\;  M_z=\sigma^5/(O/H)
\end{equation}
where $O/H$ is the metallicity. Based on the above relation and the work
of Melnick, Terlevich \& Terlevich (2000), the distance modulus of HII
galaxies can be written as:
\begin{equation}
\mu = 2.5 \log(\sigma^5/F_{H\beta})- 2.5 \log(O/H)- A_{H\beta} -26.44\;,
\end{equation}
with $F_{H\beta}$ and $A_{H\beta}$ the flux and extinction in
$H_{\beta}$, respectively. The rms dispersion
in distance modulus was found to be $\sim 0.52$ mag.
The analysis of Melnick, Terlevich \& Terlevich (2000) has
shown that most of this dispersion ($\sim 0.3$ mags)
comes from observational errors in the stellar velocity dispersion measurements, 
from photometric errors and metallicity effects. 
It is therefore possible to understand and correct the sources of
random and systematic errors of the $L(H_{\beta})-\sigma$ relation, and
indeed with the availability of new observing
techniques and instruments, we hope to reduce significantly the
previously quoted rms scatter.

A few words are also due to the possible systematic effects of the
above relation. Such effects may be related to the age of the HII
galaxy (this can be dealt with by putting a limit in the equivalent
width of the $H_{\beta}$ line, eg. $EW(H_{\beta})>$ 25 Angs; see
Melnick 2003), to extinction, to different
metallicities and environments. Also the $EW(H_{\beta})$ of HII galaxies at intermediate and
high redshifts is smaller than in local galaxies, a fact which should
also be taken into account.

We have commenced an investigation of all these
effects by using high-resolution spectroscopy of a relatively large
number of SDSS low-$z$ HII galaxies, having a
range of $H_{\beta}$ equivalent widths, luminosities, metal content and local
overdensity,  in order to understand systematics and
reduce the scatter of the HII-galaxy based distance estimator to
about half its present value, ie., our target is $\sim 0.25$ mag.
We will also define a medium and
high redshift sample (see Pettini et al. 2001; Erb et al. 2003),
consisting of a few hundred objects distributed in a range of 
high-redshifts, which will finally be used to define the high-$z$
Hubble function. 
\begin{figure}
\label{cosmoHII}
\centering
\resizebox{16cm}{7.5cm}{\includegraphics{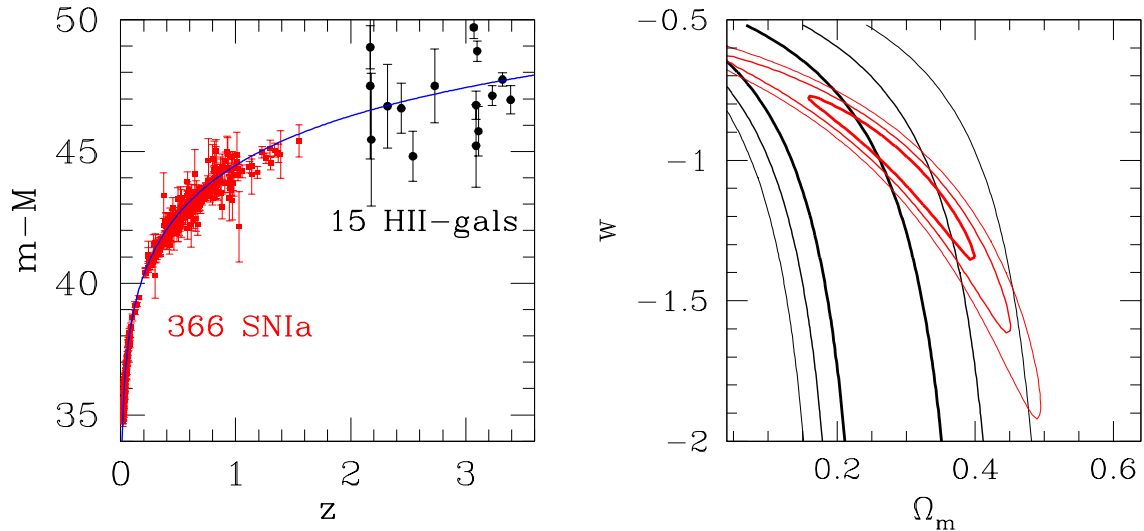}}
\caption{{\em Left Panel:} Distance moduli comparison between the {\em
    Constitution} SNIa's and the 15
  high-$z$ HII galaxies of Siegel et al. (2005). {\em Right Panel:}
The corresponding constraints in the ($\Omega_m,$ w) plane.
Although, the HII-galaxy constraints are weak, leaving completely
unconstrained the value of w, they appear to be consistent at a
$\sim 2\sigma$ level with the SNIa results. This plot serves only to
  indicate the potential of using high-$z$ HII galaxies (once of
  course we have reduced significantly their distance modulus uncertainties).}
\end{figure} 

Summarizing, the use of HII galaxies to trace the Hubble relation, as an
alternative to the traditionally used SN Ia, is based on the following
facts: 
{\em 
\begin{itemize}
\item[(a)] local and high-$z$ HII galaxies and HII regions are
physically very similar systems (Melnick et al 1987) providing a
phenomenological relation between the luminosity of the $H_{\beta}$ line, the
velocity dispersion and their metallicity as traced by $O/H$ (Melnick,
Terlevich \& Moles 1988). Therefore HII galaxies can be
used as alternative standard candles (Melnick, Terlevich \& Terlevich
2000, Melnick 2003; Siegel et al. 2005) 
\item[(b)] such galaxies can be readily observed at
much larger redshifts than those currently probed by SNIa samples, and 
\item[(c)] it is at such higher redshifts that the differences between the
predictions of the different cosmological models appear more vividly.
\end{itemize}
}

Already a sample of 15 such high-$z$ HII galaxies have been used by Siegel et al. (2005) in an
attempt to constrain
cosmological parameters but the constraints, although in the correct
direction, are very weak. 
We have performed our own re-analysis of this data-set, following however
the same procedure as that applied to the SNIa data (ie., we factorize
out the internally derived Hubble constant using a fit of the distance
moduli in the $\Omega_m, H_0$ space). The resulting
constraints on the ($\Omega_m,$w) plane can be
seen in Figure 7.  We
find that $\Omega_m\mincir 0.42$ at a 99.99\% level, independent of w.
Note that imposing w$=-1$, our analysis of the Siegel et al (2005) data set
 provides $\Omega_m=0.19\pm0.05$, which is towards the lower side of
 the generally accepted values.

Comparing these HII-based results to the present
constraints of the latest ({\em Constitution} SNIa data (right panel
of Fig. 7) clearly indicates the necessity to:
\begin{itemize}
\item re-estimate carefully the local zero-point of the
  $L(H_{\beta})-\sigma$ relation,
\item suppress the HII-galaxy distance modulus uncertainties, 
\item increase the high-$z$ HII galaxy sample by a large fraction,
\item make sure to select high-$z$ bona-fide HII-galaxies, excluding those that
  show indications of rotation (Melnick 2003). 
\end{itemize}

\section{Cosmological Parameters from the Clustering of X-ray AGN}
The method used to put
cosmological constraints, based on the angular clustering of some extragalactic
mass-tracer (Matsubara 2004, Basilakos \& Plionis 2009 and references therein),
consists in comparing the observed angular clustering with that predicted by
different primordial fluctuations power-spectra, using
Limber's integral equation (Limber 1968) 
to invert from spatial to angular clustering. By minimizing the
differences of the observed and predicted angular correlation function,
one can constrain the cosmological parameters that enter
in the power-spectrum determination as well as in Limber's inversion.
Using the latter we can relate the
angular and spatial clustering of any extragalactic population under
the assumption of power-law correlations and the small angle
approximation. 

\subsection{X-ray surveys: Biases and Systematics}
X-ray selected AGN provide a relatively unbiased census of the
AGN phenomenon, since obscured AGN, largely missed in optical surveys,
are included in such surveys.
Furthermore, they can be detected out to high redshifts and thus trace
the distant density fluctuations providing important
constraints on super-massive black hole formation,
the relation between AGN activity and Dark Matter (DM) halo hosts,
the cosmic evolution of the AGN phenomenon (eg. Mo \& White 1996,
Sheth et al. 2001), and on cosmological parameters and the dark-energy
equation of state (eg. Basilakos \& Plionis 2005; 2006).

The earlier ROSAT-based analyses
(eg. Boyle \& Mo 1993; Vikhlinin \& Forman 1995; Carrera et al. 1998;
Akylas, Georgantopoulos, Plionis, 2000; Mullis et al. 2004)
provided conflicting results on the nature and amplitude of high-$z$ AGN
clustering.
With the advent of the XMM and {\em Chandra} X-ray observatories,
many groups have attempted to settle this issue, but in vain.
Different surveys have provided
again a multitude of conflicting results, intensifying the debate
(eg. Yang et al. 2003; Manners et al. 2003;
Basilakos et al. 2004; Gilli et al. 2005; Basilakos et al 2005;
Yang et al. 2006; Puccetti et al. 2006; Miyaji et al. 2007; Gandhi et al.
2006; Carrera et al. 2007). However, the recent indications of a flux-limit dependent
clustering appears to remove most of the above inconsistencies
(Plionis et al. 2008).


Furthermore, there  are indications for a quite large high-$z$ AGN clustering length,
reaching values $\sim 15 - 18 \;h^{-1}$ Mpc at the brightest flux-limits (eg.,
Basilakos et al 2004; 2005, Puccetti et al. 2006, Plionis et al. 2008), a fact which, if verified,
has important consequences for the AGN bias evolution and therefore for the evolution of the
AGN phenomenon (eg. Miyaji et al. 2007; Basilakos, Plionis \& Ragone-Figueroa 2008).
An independent test of these results would be to establish that the environment of high-$z$
AGN is associated with large DM haloes, which being massive should be
more clustered (work in progress).

It is also important to understand and overcome the shortcomings and problems that one is facing in
order to reliably and unambiguously determine the clustering properties of the X-ray
selected AGN. Such a list of problems includes the effects of Cosmic
Variance, the so-called amplification bias, the reliability of the $\log
N-\log S$ distribution of the X-ray AGN luminosity function,
etc. (see discussion in Plionis et al. 2009).

Recently, Ebrero et al. (2009) derived the 
angular correlation function of the soft (0.5-2\,keV) 
X-ray sources using 1063 XMM-{\it Newton} 
observations at high galactic latitudes.
A full description of the data reduction, source detection and flux 
estimation are presented in Mateos et al. (2008). Note, that the 
survey contains $\sim 30000$ point sources within an effective
area of $\sim 125.5$ deg$^{2}$ (for $f_x \ge 1.4 \times
10^{-15}$ erg cm$^{-2}$ s$^{-1}$ ). 
The details regarding 
the angular correlation function, the
various biases that should be taken into account (the amplification
bias and integral constraint), the survey luminosity
and selection functions as well as issues related to possible non-AGN
contamination, which are estimated to be $\mincir 10\%$, can be found
in Ebrero et al (2009).

\subsection{Details of the Method}
An optimal approach to unambiguously determine the clustering pattern of X-ray selected AGN
would be to determine both the angular and spatial clustering pattern.
The reason being that various systematic effects or uncertainties enter differently in the
two types of analyses. On the one side, using $w(\theta)$ and its
Limber inversion, one by-passes
the effects of redshift-space distortions and uncertainties related to possible misidentification
of the optical counter-parts of X-ray sources. On the other side,
using spectroscopic or accurate photometric redshifts
 to measure $\xi(r)$ or $w_p(\theta)$ one by-passes
the inherent necessity, in Limber's inversion of $w(\theta)$, of the source
redshift-selection function (for the determination of which one uses the
integrated X-ray source
luminosity function, different models of which exist). 

The basic integral equation relating the angular and sptial correlction 
functions is:
\begin{equation}
\label{eq:angu}
w(\theta)=2\frac{H_{0}}{c} \int_{0}^{\infty} 
\left(\frac{1}{N}\frac{{\rm d}N}{{\rm d}z} \right)^{2}E(z){\rm d}z 
\int_{0}^{\infty} \xi(r,z) {\rm d}u \;\;,
\end{equation} 
where ${\rm d}N/{\rm d}z$ is the source redshift distribution,
estimated by integrating the appropriate source luminosity function
(in our case that of Ebrero et al. 2009b), folding in also the area
curve of the survey.
Note that to derive the spatial correlation length from eq. (\ref{eq:angu}), 
it is
necessary to model the spatial correlation function as a power
law, assume the small angle approximation as well as a cosmological
background model, the latter given by
$E(z)$ (from eq. 3) which for a flat background and a 
constant {\em dark energy} equation of state parameter, w, reduces to:
\be 
E(z)=[\Omega_{m}(1+z)^{3}+(1-\Omega_{m})(1+z)^{3(1+{\rm
    w})}]^{1/2}\;.
\ee
The AGN spatial correlation function is: 
\be
\xi(r,z) = (1+z)^{-(3+\epsilon)}b^{2}(z)\xi_{\rm DM}(r)\;,
\ee 
where $b(z)$ is the evolution of the linear bias
factor (eg. Mo \& White 1997; 
 Matarrese et al 1997; Sheth \& Tormen 1999; Basilakos \& Plionis 2001; 2003, Basilakos et
 al. 2008 and references therein), 
$\epsilon$ is a parameter related to the model
of AGN clustering evolution (eg. de Zotti et al. 1990)\footnote{
Following
K\'undic (1997) and Basilakos \& Plionis (2005; 2006) 
we use the
constant in comoving coordinates clustering model, ie., $\epsilon=-1.2$.}
and $\xi_{\rm DM}(r)$ 
is the corresponding correlation function of the underlying dark
matter distribution, given by the Fourier transform of the 
spatial power spectrum $P(k)$
of the matter fluctuations, linearly
extrapolated to the present epoch: 
\be
\label{eq:spat1}
\xi_{\rm DM}(r)=\frac{1}{2\pi^{2}}
\int_{0}^{\infty} k^{2}P(k) 
\frac{{\rm sin}(kr)}{kr}{\rm d}k \;\;.
\ee
The CDM power spectrum is given by: $P(k)=P_{0} k^{n}T^{2}(k)$, with
$T(k)$ the CDM transfer function 
(Bardeen et al. 1986; Sugiyama 1995)
and $n\simeq 0.96$, following the 5-year WMAP results (Komatsu et
al. 2009), and a baryonic
density of $\Omega_{\rm b} h^{2}= 0.022 (\pm 0.002)$. The
normalization of the power-spectrum, $P_{0}$, can be parametrized by
the rms mass fluctuations on $R_{8}=8 h^{-1}$Mpc scales ($\sigma_8$), 
according to:
\be
P_{0}=2\pi^{2} \sigma_{8}^{2} \left[ \int_{0}^{\infty} T^{2}(k)
 k^{n+2} W^{2}(kR_{8}){\rm d}k \right]^{-1} \;,
\ee
where 
$W(kR_{8})=3({\rm sin}kR_{8}-kR_{8}{\rm cos}kR_{8})/(kR_{8})^{3}$. 
Regarding the Hubble constant we use 
$H_{0}\simeq 71$ kms$^{-1}$Mpc$^{-1}$ (Freedman 2001; Komatsu et al. 2009).
Note, that we also utilize the non-linear corrections 
introduced by Peacock \& Dodds (1994).  

\begin{figure}
\label{bias}
\centering
\resizebox{12cm}{7.5cm}{\includegraphics{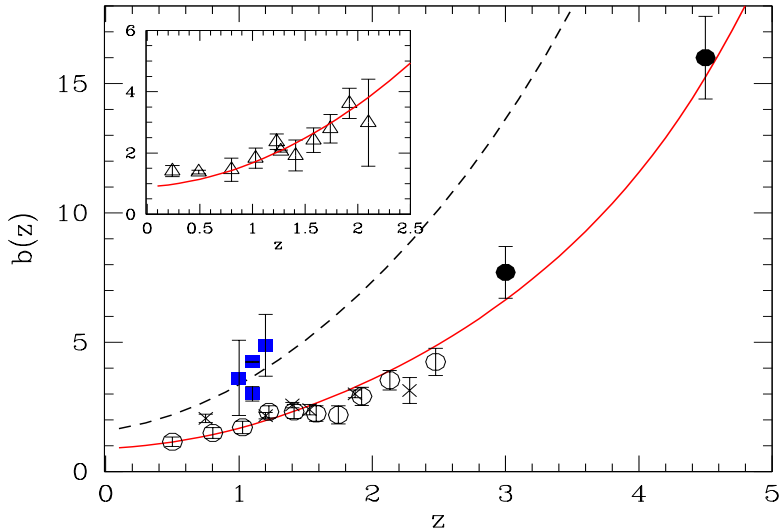}}
\caption{The observed evolution of AGN bias compared with the
  Basilakos et al. (2008) model. 
Optically selected (SDSS and 2dF) quasars (Croom et al. 2005;
Myers et al. 2007; Shen et al. 2007) correspond to circles
(empty or filled) points and crosses while 
soft-band X-ray selected AGN correspond to filled (blue) squares.
In the insert we plot the most recent 
bias values of Ross et al. (2009), based on the
SDSS optical quasar uniform sample. 
The curves reflect our bias evolution model, with the solid
(red) lines corresponding to a DM halo mass of
$M \simeq 10^{13} \;h^{-1} \; M_{\odot}$ and the dashed lines to
$M \simeq 2.5 \times 10^{13} \;h^{-1} \; M_{\odot}$.}
\end{figure}

We can now compare the observed AGN clustering
with the predicted, for different cosmological models,
correlation function of the underlying mass,
$\xi_{\rm DM}(r,z)$, and thus constrain the cosmological parameters.
To this end it is necessary to use a bias evolution model and 
although a large number of such models have been proposed in
 the literature, we use here our own approach, 
which was described initially
in Basilakos \& Plionis (2001; 2003) and extended in Basilakos,
Plionis \& Ragone-Figueroa (2008). This bias model 
is based on linear perturbation theory and the 
Friedmann-Lemaitre solutions of the cosmological field equations, while it 
also  allows for interactions and merging of the mass tracers.
Considering that each X-ray AGN is
hosted by a dark matter halo, we can analytically predict
its bias evolution behavior and conversely by comparing with
observations we can determine the
mass of the DM halo, $M_h$, within which AGN live. 

In Figure 8 we show a comparison 
between our model predictions and observationally based AGN bias results, estimated
at the sample's median redshift by:
\be
b(\bar{z})=\left(\frac{ r_{0} }{r_{0,m}} \right)^{\gamma/2} 
D^{3+\epsilon}(\bar{z})
\ee
where $r_0$ and $r_{0,m}$ are the measured AGN and the theoretical dark-matter 
clustering lengths, respectively, while $D(z)$ is the perturbation's linear 
growing mode (scaled to unity at the
present time), useful expressions of which can be found for the 
dark energy models in Silveira \& Waga (1994) and in Basilakos (2003). 

The preliminary analysis of Basilakos \& Plionis (2005; 2006) compared the measured
XMM source angular correlation function based on a shallow XMM survey,
covering a small solid angle,
with the prediction of different spatially flat and constant-w cosmological models.
The recent availability, however, of
the highly accurate
angular correlation function XMM results of Ebrero et al (2009), based on
the largest available X-ray AGN sample ($N\simeq 30000$), 
provided us the means to put stringent constraints on the cosmological parameters
(see Basilakos \& Plionis 2009).
In Table 2 (and in Figure 9) we present the
cosmological constraints provided by the current analysis.
The power of our procedure can be appreciated by comparing our constraints, 
under the prior of a flat Universe, with those provided by
most other cosmological test (see Fig.2 of Basilakos \& Plionis 2009).

\begin{table}
\caption{The best fit values from the likelihood analysis of the X-ray
  AGN clustering:
Errors of the fitted parameters represent $1\sigma$ uncertainties, while 
the parameters in bold indicate those fixed during the fitting process.}
\tabcolsep 28pt
\begin{tabular}{|cccc|} 
\hline
$\Omega_{\rm m}$& w& $\sigma_{8}$& $M_h/10^{13}h^{-1}M_{\odot}$\\
\hline 
$0.27\pm 0.04$  & $-0.90^{+0.10}_{-0.16}$ & $0.74^{+0.14}_{-0.12}$ &  $2.50^{+0.50}_{-1.50}$ \\
$0.26\pm 0.05$  & $-0.93^{+0.11}_{-0.19}$ & ${\bf 0.8}$ &  $2.0^{+0.30}_{-0.2}$ \\
$0.24\pm 0.06$  & ${\bf -1}$ & $0.83^{+0.11}_{-0.16}$&  ${\bf 2.50}$ \\ \hline
\end{tabular}
\end{table}
\begin{figure}
\label{X-SN}
\centering
\resizebox{18cm}{11cm}{\includegraphics{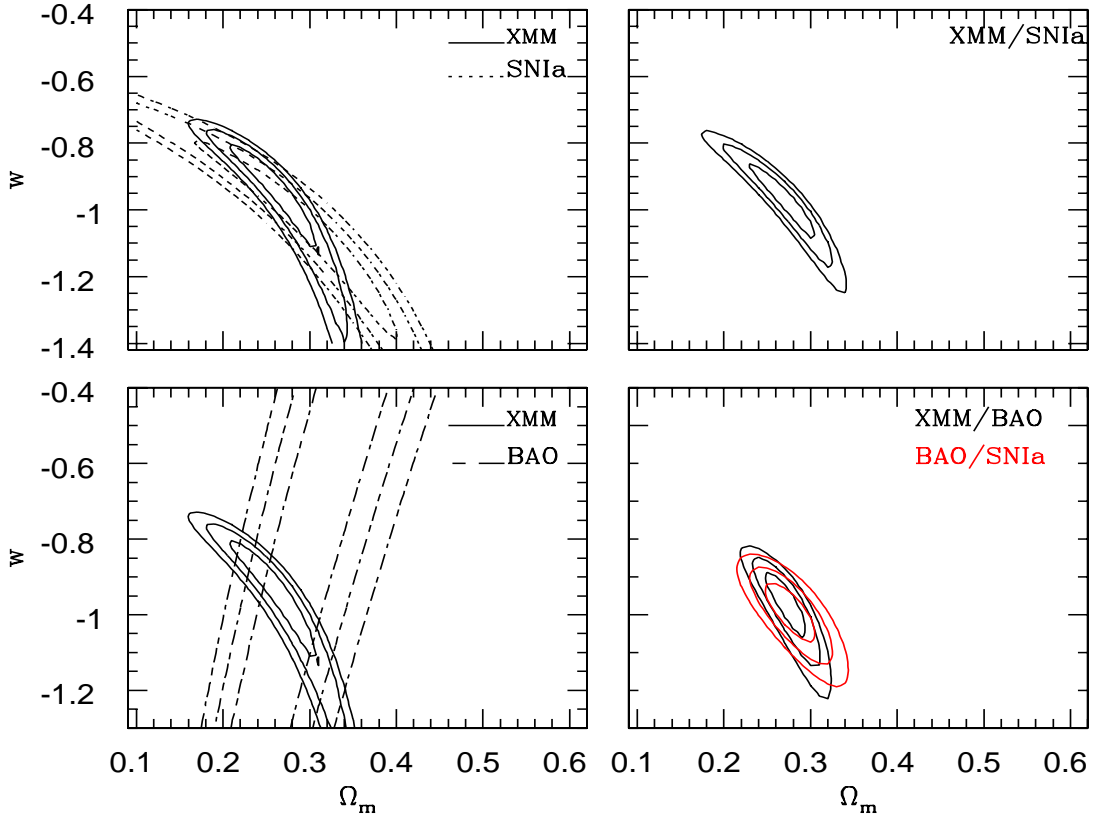}}
\caption{{\em Upper Left Panel:} Likelihood contours on the $\Omega_m,$w plane 
from the X-ray AGN clustering analysis of Basilakos \& Plionis (2009)
(solid contours) and the {\em
  Constitution} SNIa analysis (dashed contours). {\em Upper Right
  Panel:} The corresponding joint likelihood contours.
{\em Lower Left Panel:} Likelihood contours from the X-ray AGN
clustering analysis and the BAO technique (dashed contours).
{\em Lower Right Panel:} Their corresponding joint
likelihood contours (while the BAO-SNIa joint contours are shown in red).}
\end{figure}

\section{Joint Hubble-relation and Clustering analysis}
It is evident from Figure 9 (left panels) that with the new X-ray AGN
clustering analysis we have been able to reduce significantly the
degeneracy between w and $\Omega_m$ (see also Table 2). 
We can further break the degeneracy by adding the constraints provided
by either the Hubble relation technique, using here the {\em Constitution} SNIa
sample, or the baryonic acoustic oscillation technique (BAO),
which was identified by the U.S. Dark Energy Task Force as one of the four most
  promising techniques to measure the properties of the {\em dark energy}
  and the one less likely to be limited by systematic
  uncertainties. We remind the reader that BAOs are produced by pressure (acoustic) waves in the
  photon-baryon plasma in the early universe, generated by dark matter
  (DM) overdensities. At the recombination era ($z\sim 1100$),
  photons decouple from baryons and free stream while the pressure
  wave stalls. Its frozen scale, which constitutes a “standard ruler”,
  is equal to the sound horizon length, $r_s\sim 100\; h^{-1}$ Mpc
  (e.g. Eisenstein, Hu \& Tegmark 1998). This appears as a small,
  $\sim 10\%$ excess in the galaxy, cluster or AGN power spectrum (and
  its Fourier transform, the 2-point correlation
function) at a scale corresponding to $r_s$. First evidences of this excess
were recently reported in the clustering of luminous SDSS
red-galaxies (Eisenstein et al. 2005, Padmanabhan et al. 2007; 
Percival et al. 2009). 

We therefore perform a joint likelihood analysis, assuming that any
two pairs of data sets are independent (which indeed they are) and thus the
joint likelihood can be written as the product of the two individual ones.
The current joint likelihood analysis, once we impose $h=0.71$ and
$\sigma_8=0.8$ (according to Komatsu et al. 2009),
provide quite stringent and consistent constraints of the $\Omega_m$
and w parameters: 
$$\Omega_{\rm m}=0.27\pm 0.02 \;\;\;\; {\rm and}
\;\;\;\; {\rm w}=-0.96\pm 0.07\;\;\; {\rm (XMM-SNIa)}$$
$$\Omega_{\rm m}=0.27\pm 0.02 \;\;\;\; {\rm and}
\;\;\;\; {\rm w}=-0.97\pm 0.04\;\;\; {\rm (XMM-BAO)}$$
These results can be compared with the more {\em traditional} joint
analysis of the SNIa ({\em Cosntitution} set in this case) and the BAOs,
which provide:
$$\Omega_{\rm m}=0.28^{+0.02}_{-0.03} \;\;\;\; {\rm and}
\;\;\;\; {\rm w}=-0.98\pm 0.06\;\;\; {\rm (SNIa-BAO)}$$
It is evident that within the errors all the previously presented results are 
consistent, with the XMM-BAO joint analysis providing the smallest
uncertainties of the fitted parameters. This can also be appreciated by inspecting the 
lower-right panel of Figure 9, which shows that the XMM-BAO joint analysis provides 
indistinguishable constraints with the corresponding SNIa-BAO analysis.

The necessity, however, 
to impose constraints on a more general, time-evolving, {\em dark-energy} equation 
of state (eq. 3), implies that there is ample space for great
improvement and indeed our aim is to complete this project by using a
new Hubble relation analysis, based on high-$z$ HII galaxies, as detailed in these
proceedings.

\subsection*{Acknowledgments}
We thank Dr J. Ebrero for comments and for
providing us with an electronic version of the clustering results
and their XMM survey area curve. We also thank Dr. N.P. Ross for providing
us an electronic version of the QSO uniform sample bias results.
MP also acknowledges financial
support under Mexican government CONACyT grant 2005-49878.

\end{document}